\documentclass[aps,floatfix,preprintnumbers,letterpaper,ifmbe,twocolumn]{revtex4}
\usepackage{graphicx,framed}
\usepackage{epsfig}
\usepackage{color}
\usepackage{amsfonts,amsmath,amsfonts,amscd}

\begin{document}

\title{Novel mechanism for vorticity generation in black-hole accretion disks}

\author{Chinmoy Bhattacharjee$^{\,1,2}$, Rupam Das$^{\,3}$,
S.M. Mahajan$^{\,1,2,4}$}

\affiliation{$^{1}$Institute for Fusion Studies, The University of
Texas at Austin, Austin,Texas 78712, USA }

\affiliation{$^{2}$Department of Physics,The University of
Texas at Austin, Austin,Texas 78712, USA}

\affiliation{$^{3}$Department of Physical and Applied Sciences, Madonna University, Livonia, Michigan 48150, USA}
\affiliation{$^{4}$Shiv Nadar University, NH91, Tehsil Dadri, Gautam Buddha Nagar, Uttar Pradesh, 201 314, India}

\begin{abstract}
 Vorticity generation in accretion disks around Schwarzschild and Kerr black holes is investigated in the context of magnetofluid dynamics derived for both General Relativity (GR), and modified gravity formulations. In both cases, the Kerr geometry  leads to a ``stronger" generation of vorticity  than its Schwarzschild counterpart. Of the two principal sources, the relativistic drive peaks near the innermost stable circular orbit (isco), whereas the baroclinic drive dominates at larger distances. Consequences of this new relativistic vorticity source are discussed in several astrophysical settings.

\end{abstract}

\maketitle
\section{Introduction}
An exploration of the dynamics of accretion disks near compact astrophysical objects can advance our understanding of  phenomena as diverse as angular momentum transport,  jet production and gamma ray bursts \cite{shapiro2008black,vietri2008foundations,cao2013large,abramowicz2013foundations,RevModPhys.56.255,krolik1999active}. Baroclinic instability arising from misalignment between temperature and entropy gradients in a hot charged fluid rotating in an accretion disk is considered to be one of the most probable pathways for vorticity generation in astrophysics \cite{Klahr:2002sm,Petersen:2006by,Petersen:2006bw,Kulsrud:1996km}. Such vortices (electromagnetic and hydrodynamic) can be amplified through several mechanisms such as dynamos and MRI leading to a large scale vortical field\cite{hawley2002dynamical,RevModPhys.70.1,0004-637X-765-2-115}. The resulting vortical field geometry can be responsible for angular momentum transport and production, acceleration and collimation of jets in black holes, protostars, microquasars, etc. \cite{Klahr:2002sm,RevModPhys.56.255}.

In this paper, we will explore additional sources of  vorticity generation that can contribute to angular momentum transport, jet production, and collimation as well as broaden our understanding of black hole accretion, in general. Unlike the traditional ``baroclinic" mechanism, these additional drives depend on the relativistic effects-both special and general. Vorticity generation (via the relativistic drives) in the accretion disk near the Schwarzschild black hole was, previously, studied for a generalized  ``magnetofluid" in curved spacetime\cite{asenjo2013generating}. The relativistic drive for a pure barotropic system naturally emerges in the dynamics of a magnetofluid, combining kinematic and thermodynamical attributes of a hot fluid. 

Let us begin by recapturing the salient features of the magnetofluid formalism, and of earlier work on relativistic drives: 

1)The essence of the magnetofluid formalism (for a perfect fluid) lies in the construction of an antisymmetric, hybrid tensor \cite{PhysRevLett.90.035001}
\begin{equation}\label{MTensor}
M^{\mu\nu}= F^{\mu\nu} + (m/q) S^{\mu\nu},
\end{equation} 
that is a weighted sum of the electromagnetic field tensor $F^{\mu\nu}$ (weight = charge q), and the composite (kinematic-statistical) fluid tensor  $S^{\mu\nu}= \nabla^{\mu} (\mathcal{G}U^{\nu})-\nabla^{\nu}( \mathcal{G}U^{\mu})$ (weight =mass m) \cite{bekenstein1987helicity,PhysRevLett.90.035001}. The statistical factor $\mathcal{G}$ is the thermodynamic enthalpy. In terms of $M^{\mu\nu}$, the entire dynamics of the relativistic hot fluid is expressible in the succinct equation (T is the temperature of the fluid)
\begin{equation}\label{MTensorSR}
qU_{\nu}{M}^{\mu\nu}= \frac{\nabla^{\mu}p-mn\nabla^{\mu}\mathcal{G}}{n}=T\nabla^{\mu}\sigma, 
\end{equation} 
where the right-hand side is the thermodynamic force expressed in terms of the fluid entropy $\sigma$ using the standard thermodynamic relation between entropy with enthalpy. Here, $U^{\mu}$ and $n$ represent, respectively, the plasma  4-velocity and the number density.

2) The 3-vector part of (\ref{MTensorSR}) reduces to the more familiar  form of 3D vortex dynamics except that the  standard  fluid vorticity is replaced by the  hybrid magnetofluid vorticity. The addition of relativity, however, introduces a fundamental change; the topological helicity invariant $\mathcal{H} = <\vec{\Omega}\cdot {\nabla\times}^{-1}\vec{\Omega}>$ (with $\vec{\nabla}\times^{-1}\vec{\Omega}$ being the inverse curl of vorticity) of an ideal nonrelativistic fluid no longer pertains. 

Through the ``distortion" of space-time, relativistic dynamics breaks the helicity invariant even in ideal dynamics ($\sigma=\sigma(T))$; new sources and sinks appear and the creation and destruction of the generalized vorticity become possible in ideal dynamics \cite{PhysRevLett.90.035001,PhysRevLett.105.095005,Mahajan_Yoshida,asenjo2013generating}. Such sources can, therefore, be available to  create vorticity in the accretion disk. 

3) Recently, this formalism was generalized to incorporate nonminimal coupling of the magnetofluid to a curved background space-time \cite{PhysRevD.91.064055}. It is quite remarkable, that the nonminimal coupling [introduced through $f_m(R)$, a function of  the Ricci scalar $R$] changes the equation of motion (\ref{MTensorSR}) only minimally 
\begin{equation}\label{MTensorG}
qU_{\nu}\mathcal{M}^{\mu\nu}=  Q T\nabla^{\mu}\sigma; 
\end{equation} 
it multiplies the right-hand side with a curvature dependent factor $Q=(1+\lambda f_m(R))$ that reduces to unity as the nonminimal part goes to  zero, as expected. This has also resulted in the introduction of additional gravity-coupled flow field tensor which, after appropriate 3+1 decomposition, yields new expressions for generalized electric and magnetic field.  
The formalism, epitomized in Eqs.  (\ref{MTensorSR})-(\ref{MTensorG}), will be, henceforth, called Magnetofluid formalism.

In this paper, then, we will investigate vorticity generation in accretion disk for minimally as well as nonminimally coupled magnetofluid in Schwarzschild and Kerr space-time.  The magnitude of  induced vorticity will be estimated in the special case  when matter and space-time are coupled with constant Ricci scalar $R_0$, the simplest functional form of nonminimal coupling. 

We first give a summary of the derivation of generalized equation of motion of a new hybrid magnetofluid in curved background space-time.  Next, the Arnowitt-Deser-Misner (ADM) formalism of electrodynamics \cite{MTW,thorne1982electrodynamics,Wald,thorne1986black} presented in Appendix \ref{magnetofluiddynamics} is applied to this new formulation of magnetofluid and the equations obtained are cast into the vorticity evolution equation.  These equations are analyzed for accretion disks to calculate and estimate  generalized vorticity. Finally we compare the  relativistic drives  with the more conventional baroclinic drive.

\section{Magnetofluid Formalism}\label{MFformalism}

The modified theory of gravity offers an alternative approach to explain the inferred accelerated expansion of the Universe and other cosmological data by  introducing deviations from the GR. The so-called $f(R)$ gravity modifies GR in the low energy (curvature) regime, but its behavior in the high energy regime has also been the topic of current research  \cite{chang2011stellar,PhysRevD.85.123006,2013A&A551A4P}. To expand the scope of our earlier calculations, we incorporate the $f(R)$ gravity [through the term $\lambda f(R)$]  in the magnetofluid formalism. 
Referring the reader to Ref \cite{PhysRevD.91.064055} for a detailed derivation, we simply write down, here, the two main equations: the modified Einstein equation ($G = c = 1$)

\begin{align}\label{modified_eintsein}
(1+F_{g}(R)) R^{\mu\nu} - \frac{1}{2} (R+f_{g}(R)) g^{\mu\nu} \nonumber\\
-(\nabla^{\mu}\nabla^{\nu}-g^{\mu\nu})F_{g}(R)= 8\pi T^{\mu\nu}_{total},
\end{align}
and the magnetofluid equation of motion
\begin{align}\label{eom1}
(1+\lambda f_{m}(R))\nabla_{\mu}T^{\mu\nu}_{pf}= \nonumber\\
\left[q n F^{\nu} \ _{\beta}U^{\beta} 
-\lambda F_{m}(R)(T^{\mu\nu}_{pf}+g^{\mu\nu}\rho)\nabla_{\mu}R\right],
\end{align}
where $F_{g}=f_{g}'(R)$, $R$ is the Ricci scalar and  $ T^{\mu\nu}_{total}$ is the total stress-energy tensor for both the perfect fluid and Maxwell's field in curved space-time.  In the preceding equations,  $\tau$ is the proper time, $q$ is the charge of the particle, $R^{\mu\nu\alpha\beta}$ is the Riemann tensor and $ T^{\mu\nu}_{pf}=(p+\rho)U^{\mu}U^{\nu}+pg^{\mu\nu}$ (with $U^{\mu}=dx^{\mu}/{d\tau}$) is the energy-momentum tensor for a perfect isotropic fluid.  The phenomenological parameter $\lambda$  represents the coupling strength of the plasma to its background geometry, now modified through $f_{m}(R)$ and $F_{m}(R)=f_{m}'(R)$. The quantity $p+\rho=h$ is the enthalpy of the fluid plasma and often appears in the formalism as the combination $\mathcal{G}=h/mn$ with $m$, $n$, $\rho$, and $p$ being the mass, number density, energy density, and pressure, respectively.  \\


Notice that the equation of motion (\ref{eom1}) is not yet in the promised ``canonical" form (\ref{MTensorG}). To make progress, following Refs. \cite{PhysRevLett.90.035001,PhysRevD.91.064055}, we will construct the new grand unified vorticity tensor $\mathcal{M}^{\nu\mu}$ that reflects nonminimal coupling.  After some patient algebra, we find that
$\mathcal{M}^{\nu\mu}$ is again the weighted sum as [\ref{MTensorSR}],

\begin{equation}\label{unifiedtensordef}
\mathcal{M}^{\mu\nu}=F^{\mu\nu}+\frac{m}{q}D^{\mu\nu}
\end{equation}
but with a considerably more complicated 
\begin{equation}\label{unifiedtensordef}
D^{\mu\nu}=(1+\lambda f_{m}(R) -\lambda RF_m)S^{\mu\nu}+\frac{m}{q}\lambda F_mK^{\mu\nu}
\end{equation}
replacing $S^{\mu\nu}=\nabla^{\mu}(\mathcal{G} U^{\nu})-\nabla^{\nu}(\mathcal{G}U^{\mu})$. We needed to ``find"  a new curvature-weighted antisymmetric flow field tensor 
\begin{equation}\label {CurvFluidTen}
K^{\mu\nu}=\nabla^{\mu}(R\mathcal{G}U^{\nu})-\nabla^{\nu}(R\mathcal{G}U^{\mu}),
\end{equation}
to derive the sought-after form. The new fluid tensor $D^{\mu\nu}$ contains, explicitly, the coupling of flow field to gravity. Thus, the dynamics of a hot fluid system in curved background space-time can be written into the canonical four dimensional vortex form 
\begin{equation}\label{finaleom}
qU_{\nu}\mathcal{M}^{\mu\nu}=(1+\lambda f_{m}(R))T\nabla^{\mu}\sigma,
\end{equation}
the form advertised in (\ref{MTensorG}).
We have ``assumed" that the standard thermodynamic relations continue to hold; it is, of course,
contingent  upon an appropriately well-defined local concept of temperature in curved space-time. \\

Notice that,  when $\lambda = 0$, $\mathcal{M}^{\mu\nu}$ reduces to its minimally coupled  counterpart tensor $M^{\mu\nu}$ defined in [\cite{PhysRevLett.90.035001,asenjo2013generating}]. \\

Equation (\ref{finaleom}) is the main result  that describes the magneto fluid dynamics in curved space-time. It reveals that a charged relativistic fluid, coupled nonminimally to gravity, obeys a 4D vortex dynamics like its gravity free and minimally coupled (to gravity) counterparts. The new grand vorticity tensor subsumes earlier limiting cases in a  transparent manner.

We would now apply the above formulation to investigate the vorticity generation in accretion disks around black holes. To do calculations in terms of familiar quantities, an appropriate 3+1 decomposition of the spacetime is necessary; it is presented in Appendix \ref{magnetofluiddynamics}. Next, we present the 3D vortical dynamics in order to facilitate computation of vorticity generation.

\subsection{Vortical dynamics}\label{vorticialdynamics}
With the 3+1 decomposition presented in Appendix \ref{magnetofluiddynamics}, the spacelike projection, i.e., $\gamma^{\beta}\ _{\mu}$ projection of the unified field equation of motion (\ref{finaleom}) gives us the momentum evolution equation
\begin{equation}\label{generalmomeq}
\alpha q \Gamma \vec{\xi}+q\Gamma(\vec{v}\times\vec{\Omega})=-(1+\lambda f_{m}(R))T\vec{\nabla}\sigma
\end{equation}
whereas the timelike ( $n_{\mu}$) projection gives the equation of energy conservation
\begin{equation}\label{generalenergyeq}
\alpha q\Gamma \vec{v}\cdot\vec{\xi}=T (1+\lambda f_{m}(R))(\mathcal{L}_t\sigma-\vec{\beta}\cdot\vec{\nabla}\sigma),
\end{equation}
where $\vec{\xi}$ and $\vec{\Omega}$ are, respectively, the generalized electric and magnetic fields given by Eqs. (\ref{generalefield}) and (\ref{generalbfield}) in Appendix \ref{magnetofluiddynamics}. Also, $\alpha$, $\Gamma$, and $\vec{v}$ are defined in Eq. (\ref{canonicalmetric}), (\ref{lorentzfactor}), and (\ref{fourvelocity}).
Sources responsible for magnetic field generation, in particular, the sources that are gravity driven, can be derived from the generalized vorticity evolution equation (which is really the generalized Faraday's law) by manipulating Eq. (\ref{generalmomeq}). 

Since $\mathcal{M}^{\mu\nu}$ is an antisymmetric tensor, the divergence of its dual is zero, i.e., $\nabla_{\mu}\mathcal{M}^{*\mu\nu}=0$. Taking  the $\gamma^{\beta}\ _{\mu}$  projection of the preceding identity, we derive
\begin{equation}\label{faradaygeneral}
\mathcal{L}_t\vec{\Omega}=\mathcal{L}_{\vec{\beta}}\vec{\Omega}-\vec{\nabla}\times(\alpha\vec{\xi})-\alpha \Theta \vec{\Omega},
\end{equation}
where $\mathcal{L}$  denotes Lie derivatives with $\mathcal{L}_t = \partial_t$ along $t^{\mu}$,  $\mathcal{L}_{\vec{\beta}}{\vec{\Omega}}= [\vec{\beta},\vec{\Omega}]$, and the expansion factor $\Theta$ is defined in Appendix \ref{magnetofluiddynamics}.

It should be noted that, even in the absence of nonminimal coupling to gravity ($\lambda=0$), (minimal) coupling to gravity still manifests in the formalism. 
Equation (\ref{faradaygeneral}), in conjunction with Eq.  (\ref{generalmomeq}), gives us the vorticity evolution equation 
\begin{equation}\label{vorticity}
\mathcal{L}_t\vec{\Omega}-\vec{\nabla} \times (\vec{v}\times \vec{\Omega})-\mathcal{L}_{\vec{\beta}}\vec{\Omega}+\alpha \Theta \vec{\Omega}=\vec{\nabla}\times\left(\frac{T}{q\Gamma}(1+\lambda f_{m}(R))\vec{\nabla}\sigma\right).
\end{equation}

All terms on the left-hand side  operate on the vorticity 3-vector $\vec{\Omega}$ while the right-hand side  provides, just as in the conventional picture, possible sources for vorticity generation.  
The left-hand side, however, has  lot more structure than the conventional 3D vortex dynamics; the first two terms reflect the standard Helmholtz vortical dynamics, 
while $\alpha \Theta \vec{\Omega}$ and $\mathcal{L}_{\vec{\beta}}\vec{\Omega}$, are nontrivial gravity modifications. Thus, the gravity coupling does, fundamentally, modify the projected 3D vortex dynamics, in spite of the fact that the 4D vortex equations had exactly the same form.
\section{Vorticity generation}\label{vorticitygeneration}

 To apply the formalism to vorticity generation in astrophysics, specifically in accretion disks around compact objects like Schwarzschild and Kerr black holes, we have to specify the space-time geometry the space-time metric that controls the motion of plasma particles. The standard metric describing the stationary and axially symmetric (or spherically symmetric) spacetime for Kerr (or Schwarzschild) black holes can be written as \cite{Harkolobo}
\begin{equation}\label{ASmetric}
ds^2=g_{tt}dt^2+2g_{t\phi}dtd\phi+g_{rr}dr^2+g_{\theta\theta}d\theta^2+g_{\phi\phi}d\phi^2.
\end{equation}
The exploration of geodesic motions of plasma in accretion disks will allow us to compute various relevant physical quantities. Since we are interested only in the timelike geodesics in thin accretion disks,  the Euler-Lagrange equations can be derived from the Lagrangian for the above stationary and axisymmetric spacetime, $
2L = ds^{2}/d\tau ^{2} = -1$,
with $\tau$ being the proper time along timelike geodesics. Thus, the corresponding Euler-Lagrangian equations describing the timelike geodesics in the equatorial plane take the form (\cite{Harkolobo})
\begin{equation}\label{ELequation1}
\frac{dt}{d\tau}=\frac{\tilde{E}g_{\phi\phi}+\tilde{L}g_{t\phi}}{g_{t\phi}^2-g_{tt}g_{\phi\phi}},
\end{equation}
\begin{equation}\label{ELequation2}
\frac{d\phi}{d\tau}=-\frac{\tilde{E}g_{t\phi}+\tilde{L}g_{tt}}{g_{t\phi}^2-g_{tt}g_{\phi\phi}}, 
\end{equation}
\begin{equation}\label{ELequation3}
g_{rr}\left(\frac{dr}{d\tau}\right)^2=-1+\frac{\tilde{E}^{2}g_{\phi\phi}+2\tilde{E}\tilde{L}g_{t\phi}+\tilde{L}^{2}g_{tt}}{g_{t\phi}^2-g_{tt}g_{\phi\phi}}\equiv V_{eff},	
\end{equation}
where  $\tilde{E}$ and $\tilde{L}$ are specific energy and specific angular momentum respectively.
For stable circular orbits in the equatorial plane, using $V_{eff}=0$ and $dV_{eff}/dr=0$, the constants of motion including angular velocity $\omega$ are found to be
\begin{eqnarray}\label{definition}
\tilde{E}=-\frac{g_{tt}+g_{t\phi}\omega}{\sqrt{-g_{tt}-2g_{t\phi}\omega-g_{\phi\phi}\omega^2}},\\
\tilde{L}=\frac{g_{t\phi}+g_{\phi\phi}\omega}{\sqrt{-g_{tt}-2g_{t\phi}\omega-g_{\phi\phi}\omega^2}},\\
\omega=\frac{d\phi}{dt}=\frac{-g_{t\phi,r}+\sqrt{(g_{t\phi,r})^2-g_{tt,r}g_{\phi\phi,r}}}{g_{\phi\phi,r}}.
\end{eqnarray}
The Lorentz factor for particles can be derived from
\begin{equation}\label{gammadef}
\Gamma=\frac{\tilde{E}g_{\phi\phi}+\tilde{L}g_{t\phi}}{g_{t\phi}^2-g_{tt}g_{\phi\phi}}=\frac{1}{\sqrt{-g_{tt}-2g_{t\phi}\omega-g_{\phi\phi}\omega^2}}.
\end{equation} 
Moreover, for any given scalar function $P$, the gradient is defined as 
\begin{equation}\label{gradscalar1}
\vec{\nabla} P=\frac{1}{\sqrt{g_{rr}}}\partial_rP \ \hat{e}_r+\frac{1}{\sqrt{g_{\theta\theta}}}\partial_{\theta}P\ \hat{e}_{\theta}+\frac{1}{\sqrt{g_{\phi\phi}}}\partial_{\phi}P \ \hat{e}_{\phi}.
\end{equation}

Next, we assume a thin accretion disk with zero latitudinal speed $v_{\theta}=0$ for the plasma;  we will also assume  that the radial velocity of the plasma is negligible compared to the orbital velocity $v_{\phi}>>v_r$. The orbits of the plasma constituents are also taken to be almost circular ($\dot{r}\approx 0$). Since our formalism is based on perfect fluid, we can also assume the plasma to be barotropic with its pressure depending on density only, i.e., $\sigma=F(T)$. Then, with this assumption, we can write the relation between the temperature and entropy gradient as $T\vec{\nabla}\sigma= \chi k_b\vec{\nabla} T$, which evidently will cause the baroclinic drive $N_B$ to vanish, where $\chi$ is a dimensionless quantity of order unity.

\begin{figure}[ht]
	\includegraphics[width=0.45\textwidth]{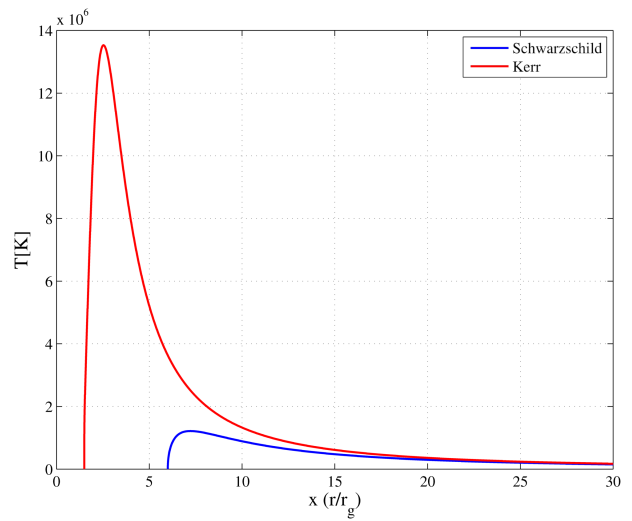}
	\caption{Temperature profile for the Schwarzschild (blue) and Kerr black hole (red) from $x=1.5$ to $x=30$. }
	\label{fig:temp}
\end{figure}

\begin{figure}[ht]
	\includegraphics[width=0.45\textwidth]{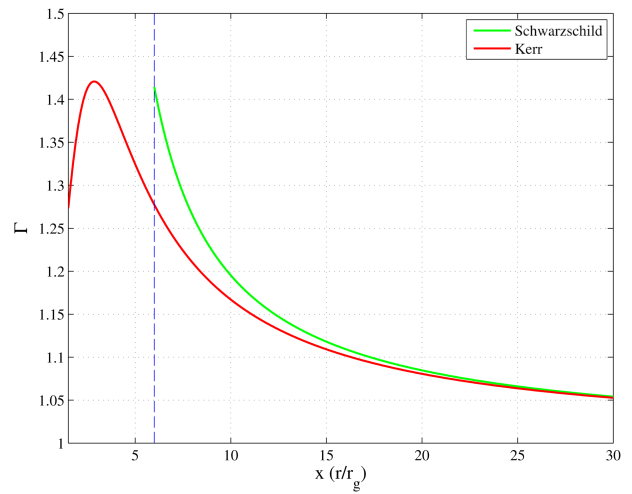}
	\caption{ Lorentz factor for Schwarzschild (green) and Kerr black hole (red) from $x=1.5$ to $x=30$. The dashed line (blue) indicates inner-most circular orbit (isco) for the Schwarzschild blackhole. }
	\label{fig:gamma}
\end{figure}

To compute the appropriate temperature profile in the region of interest in the accretion disk, we follow the prescription presented by Novikov and Thorne \cite{novikov1973astrophysics}.  It turns out that, for $M=14.3 M_{\odot}$ and $\dot{M}=0.472\times 10^{19}g s^{-1}$, the inner- and outermost stable circular orbits (average width of the accretion disk) are located mostly in the optically thick region of the accretion disk ; this is true for both geometries. Therefore, for an optically thick region, using the Stefan-Boltzmann law, the temperature profile can be written as 
\cite{Harkolobo,2013A&A551A4P}
\begin{equation}
\label{blackbody}
T(r)=z\left(\frac{\mathfrak{f}(r)}{\sigma_{SB}}\right)^{\frac{1}{4}},
\end{equation}
where $\sigma_{SB}$ is the Stefan-Boltzmann constant and $z$ is redshift due to gravitational effects. For $\theta=\pi/2$, and for a vanishing disk inclination angle, the redshift can be written as $1+z=\Gamma$. Here $\mathfrak{f}(r)$ is the energy flux for a relativistic accretion disk presented in Refs. [\cite{page1974disk,2013A&A551A4P}] by Page and Thorne as 
\begin{equation}
\mathfrak{f}(r)=-\frac{\dot{M}_0}{4\pi\sqrt{-g}}\frac{\omega,_{r}}{(\tilde{E}-\omega\tilde{L})^2}\int^{r}_{r_{isco}}(\tilde{E}-\omega\tilde{L})\tilde{L},_{r}dr,
\end{equation}
where $r_{isco}$ is the radius of the innermost stable circular orbit in the accretion disk, and $\dot{M}_0$ is the mass accretion rate. The temperature and the Lorentz factor profiles, displayed, respectively, in Figs $(\ref{fig:temp})$ and $(\ref{fig:gamma})$, reveal similar general features for the Kerr space-time: increasing from their corresponding value at $r_{isco}$, they  reach a peak at some radius and then monotonically decrease as we move away from the center of the corresponding black holes \cite{skadowski2011relativistic}. However, only the temperature profile in the Schwarzschild geometry shares the similar feature. These features of the temperature and the gamma ($\Gamma$) profiles will manifest in the vorticity generation as well as in the relative strength between the corresponding relativistic and classical drives.

Throughout this paper, we will use  parameters obtained from observation on  the Galactic black hole Cygnus-XI as representative: $M=14.38 M_{\odot}$, $\dot{M}=0.472\times 10^{19}g s^{-1}$ and a$=0.99$,  where $M$ and $a$ are, respectively, the mass, and angular momentum per unit mass of the black hole \cite{orosz2011mass,gou2011}.
Figure $\ref{fig:temp}$ shows the temperature profile in Kerr (Schwarzschild)  geometry from $r=1.5r_g$ ($r=6r_g$) to $r=30r_g$ with $r_g=GM/c^2$ in the normalized unit of $x=r/r_g$. The profile shows a peak temperature between $10^6-10^7K$ which drops as we move from the event horizon radially outward. These profiles are used in this paper to calculate the vorticity generation in the accretion disk.

\subsection{Schwarzschild geometry}

For a spherically symmetric and static space-time (Schwarzschild space-time), the above vortical evolution equation (\ref{vorticity}) reduces to the one presented in Ref. \cite{asenjo2013generating}, i.e., $\mathcal{L}_t\vec{\Omega}-\vec{\nabla} \times (\vec{v}\times \vec{\Omega})=\vec{\nabla}\times\left((T/q\Gamma)\vec{\nabla}\sigma\right)$. Since the spherically symmetric and static space-time can be foliated without any shift function $\vec{\beta}$, and the foliation obeys the time translation symmetry leading to a vanishing extrinsic curvature, the new term involving $\Theta$ on the left hand side disappears. Thus, the structure is precisely like  the 3D vortex dynamics. The simplified vortical evolution equation can be used to approximately compute the weak field seed generation in  the hot fluid system in the accretion disk in Schwarzschild geometry.

The relevant space-time metric elements are 

\begin{alignat}{2}\label{smetric}
g_{tt}=-\alpha^2=-(1-\frac{r_0}{r})\ ;\  &\qquad\text{} g_{rr}=(1-\frac{r_0}{r})^{-1},\notag \\
g_{\theta\theta}=r^2\ ;\  &\qquad\text{}  g_{\phi\phi}=r^2\sin^2\theta.
\end{alignat}

Then, using the Eqs. (\ref{definition}-\ref{gammadef}), we can calculate the orbital velocity and gamma factor for the orbital motion of plasma in the accretion disk.
Since the radial velocity of the plasma is assumed to be negligible compared to the orbital velocity $v_{\phi}>>v_r$, the orbits of the plasma elements are taken to be almost circular ($\dot{r}\approx 0$). 
Then, inserting the Schwarzschild metric elements, (\ref{smetric}),  in Eqs. (\ref{ELequation1}-\ref{ELequation3}) and imposing  $V_{eff}=0$ and $dV_{eff}/dr=0$ reveals that there exists one stable circular orbit at $x > 6$ and one unstable circular orbit at $6>x>3$  [\cite{Wald}]. This also dictates the applicability of the temperature profile in accretion disk. The temperature profile used in this paper is valid in the region of stable circular orbits.

\subsection{Kerr geometry}
For an axisymmetric (but not spherically symmetric) stationary system, like the Kerr black hole, our previous assumption of the zero shift function, $\vec{\beta}=0$, is no longer valid. Consequently, the pertinent  equation (\ref{vorticity}), in general,  does not show any similarity to standard 3D vortex dynamics. The shift function for rotating black holes can be taken to be $\vec{\beta}=-\omega \ \hat{e}_{\phi}$ with respect to a zero angular momentum observer. The term involving the shift function, however,  will give zero contribution since we assume that both $\omega$ and $\Omega$  have only radial dependence. In addition, it can be further shown that the term involving the expansion factor $\Theta$ vanishes. Thus, the vorticity evolution equation, even for Kerr geometry,  will resemble  the standard 3D vortex dynamics.\\  

The relevant space-time metric elements in Boyer-Lindquist coordinates are \cite{2013A&A551A4P}
\begin{alignat}{2}\label{kerrmetric1}
g_{tt}=-\alpha^2=-\frac{(\Delta_r-a^2)}{r^2}\ ;\  &\qquad\text{} g_{rr}=\frac{r^2}{\Delta_r},\notag \\
g_{t\phi}=-\frac{2a}{r^2}(r^2+a^2-\Delta_r)\ ;\  &\qquad\text{} g_{\phi\phi}=\frac{(r^2+a^2)^2-\Delta_ra^2}{r^2},
\end{alignat}
where $\Delta_r=(r^2+a^2)-2r_gr$. Then, again inserting the Kerr metric elements, (\ref{kerrmetric1}),  in Eqs. (\ref{ELequation1}-\ref{ELequation3}) and imposing  $V_{eff}=0$ and $dV_{eff}/dr=0$ reveals  that the inner most circular orbit in the Kerr black hole is located at $x=1.4545$ for $a=0.99r_g$, which was taken into account in deriving the temperature profile for the accretion disk in Kerr geometry. Note that we will assume $a=0.99r_g$ throughout the rest of the paper.

\subsection{Computing vorticity }
For both Schwarzschild and Kerr configurations, computation of vorticity generation requires knowledge of the $\phi$ dependence of the temperature profile. However, the most commonly used temperature profiles (including the GR corrected ones) for accretion disks show only radial dependence. Previously, an estimate of vorticity generation was computed using an average temperature of the accretion disk \cite{asenjo2013generating}. However, as shown in Fig. (\ref{fig:temp}), General Relativity restricts the application of an average disk temperature throughout the disk as it involves regions of unstable orbits leading to nonlinear behavior. Moreover, plasmas orbiting in accretion disks for both black hole configurations undergo gravitational radiation reaction, which for a Kerr black hole can cause a plasma particle to lose as much as 42 percent of its initial energy as it approaches the event horizon \cite{Wald}. 

Toroidal temperature dependence is created due to the  gravitational radiation by the orbiting plasma particles; the induced radiation reaction, in turn,  makes the stable circular orbits deviate slightly from geodesic motion  \cite{Wald}. A particle, initially in a circular orbit at $x > 6$ ($x>1.4545$) for the Schwarzschild (Kerr) metric, slowly spirals into smaller nearly circular orbits as it radiates energy until it reaches the orbital radius $x=3$ ($x=1.4545$), where the orbit becomes unstable. Therefore, the stable circular orbits do not close in either  geometry, and, depending on the magnitude of radiation reaction, the spatial orbital trajectory in the equatorial plane is assumed to be represented by $r=r(\phi)$,  a solution to the geodesic equation relating coordinates $r$ to $\phi$. 

Next, we assume, without loss of generality, that the orbits of the spiraling plasma elements in the disk can be approximated by $r(\phi) = r_{0}e^{-\zeta \phi}$, where $r_{0}$ is the initial radial distance of the plasma particles from the center of the black hole, and $\zeta$ is the parameter that controls how tightly a nearly  circular orbit spirals around the black hole.  In general, the factor $\zeta$ can be a complicated function of black hole mass as well as the energy and the angular momentum of plasma elements. Determined from the geodesic equation relating coordinates $r$ to $\phi$ with appropriate boundary conditions, $\zeta$  can be a function of the radial distance. However, since our focus is on the region of accretion disk over which the timelike geodesic orbits of the plasma elements are closely bound, we can assume the spiraling parameter $\zeta$ to be a constant. A rapidly varying spiral with varying $\zeta$ will contribute more to the vorticity as can be seen from Eq.(\ref{generic1}). 

As mentioned earlier, the radiation reaction, can be the source  for  slight deviations from the closed stable circular orbit, thereby imparting  temperature variations  around a spirally circular orbit. Thus, for the spiral orbit, $r(\phi)= r_{0}e^{-\zeta \phi}$, $\partial T / \partial \phi=(\partial T/\partial r)({d r}/{d \phi})= -\zeta r (\partial T/\partial r)$, and the seed vortical field $|\vec{\Omega}|$ may be estimated as
\begin{align}\label{generic1}
\vec{\Omega}(r)=-\frac{\chi k_b c r\zeta}{q}\frac{1}{\sqrt{g_{rr}g_{\phi\phi}}}\partial_r(\Gamma^{-1})(\partial_rT) \Delta t \ \hat{\theta},
\end{align}
where $\Delta t$ is the characteristic time for linear vorticity generation under which the changes in space-time geometry are negligible. Therefore, we choose this time-scale to be $\Delta t=2\pi/\omega$, and, as expected, this coordinate time interval $\Delta t$ is related to the proper time interval $\Delta \tau$ by $\Delta t = \Gamma \Delta \tau$. For an observer far away from the accretion disk under observation, these two time intervals are practically the same.  For the Schwarzschild geometry, Eq. (\ref{generic1}) simplifies  to
\begin{align}\label{genericsch}
\vec{\Omega}(x)=-\frac{3\zeta k_b\pi\Gamma\alpha\chi}{q\sqrt{x}r_g}\partial_xT \ \hat{\theta},
\end{align}
$x=r/r_g$ is the normalized distance, and $q=-e$ is the electron charge.

\begin{figure}
	\includegraphics[width=0.90\linewidth]{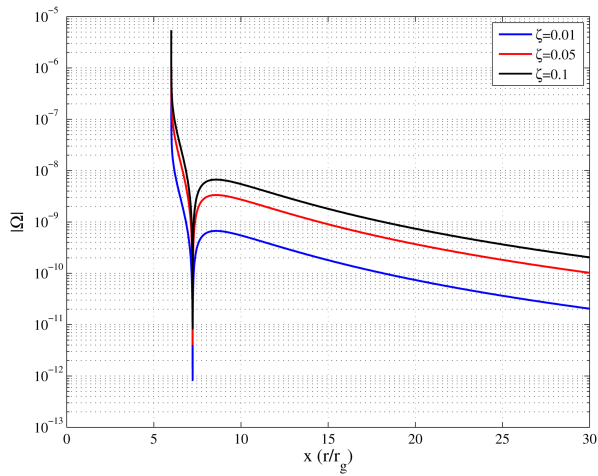}
	\caption{Vorticity magnitude (in the units of Gauss) for the Schwarzschild black hole with different $\zeta$ values between $6.0 r_g$ and $30 r_g$ for  $M=14.8M_{\odot}$}
	\label{fig:sbh_reg}
\end{figure}

\begin{figure}
	\includegraphics[width=0.90\linewidth]{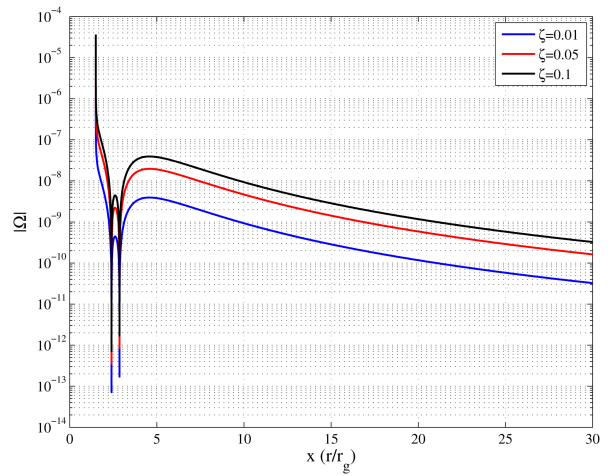}
	\caption{Vorticity magnitude (in the units of Gauss)for the Kerr black hole with different $\zeta$ values between $1.5 r_g$ and $30 r_g$ for  $M=14.8M_{\odot}$ with an accretion rate  $\dot{M}=0.472\times 10^{19}g/s$ and a=0.99.}
	\label{fig:kbh_reg}
\end{figure}

Figure \ref{fig:sbh_reg} and \ref{fig:kbh_reg} show the radial dependence of the magnitude of the vorticity generated  in  the accretion disk plasmas embedded in Schwarzschild and Kerr  geometries  for three different choices of $\zeta$. Both figures show a drastic reduction in $|\Omega|$ as the temperature maxima  are approached (temperature gradient going to zero), and then $|\Omega|$ picks up as we go over to the other side of the \textbf{maximum.} In addition, the existence of a second dip in the Kerr vorticity profile can be attributed to the vanishing of the gradient of the corresponding Lorentz factor. Therefore, the effect of the relevant gamma and temperature gradients indicates a distinct vorticity profile in Kerr space-time, where vorticity changes direction twice before it gradually decays radially outward in the disk.

However, as we move radially outward in the disk, both induced vorticities decrease almost at the same rate maintaining their difference.

\subsection{Black hole accretion disk in modified gravity}

The results of last section were derived for the accretion disk fluid minimally coupled to standard GR. We will, now explore the changes in vorticity generation brought about by modified gravity, more precisely,  by  the simplest functional form of $f(R)=R_0 = constant$.  Following the discussion presented in Sec. \ref{vorticitygeneration}, plasma particles in the Schwarzschild accretion disk are described by following metric components
\begin{alignat}{2}\label{modsmetric}
g_{tt}=-\left(1-\frac{2r_g}{r}-\frac{R_0}{12}r^2\right)=\alpha^2\ ;  &\qquad\text{} g_{rr}=\frac{1}{\alpha^2},\notag \\
g_{\theta\theta}=r^2\ ;\  &\qquad\text{}  g_{\phi\phi}=r^2\sin^2\theta,
\end{alignat}
where  $R_0$ is the constant Ricci scalar. First, setting $g_{tt}(r)=0$, we obtain the condition on Ricci scalar  $R_0$
\begin{equation}\label{sceq}
\mathcal{R}_{0}x^3-12x+24=0
\end{equation}
with $\mathcal{R}_{0}=R_{0}r_{g}^2$ and $x=r/r_g$.
 To get a black hole without any naked singularity and with event/cosmological horizons, the solution of Eq.(\ref{sceq}) restricts the value for $\mathcal{R}_{0}$ to be $(-\infty , 4/9)$  (\cite{2013A&A551A4P}). Further restrictions on $\mathcal{R}_{0}$ to isolate the geodesics for stable circular orbits can be imposed by demanding $V_{eff}=0$, $dV_{eff}/dr=0$, and $d^{2}V_{eff}/d^{2}r\geq 0$,
\begin{align} \label{r0}
\mathcal{R}_{0}=\frac{12(6-x_c)}{(15-4x_c)x^3_{c}}
\end{align}
with $x_c$ being the radius of the stable circular orbit. The above relation also reveals that there exist one innermost and one outermost stable circular orbit, and the upper limit on $\mathcal{R}_{0}$ reduces to $2.85 \times 10^{-3}$. However, the choice of an exact value of $\mathcal{R}_{0}$ befitting our current analysis depends on the average width, the temperature, and luminosity profiles of black holes. Thus, if the temperature and luminosity profiles prescribed by Page and Thorne \cite{page1974disk} are taken into account, it turns out that, for $f(R)$ Schwarzschild black holes, the new range of $\mathcal{R}_{0}$ further reduces to $(-\infty , 10^{-6}]$. Then the only judicious choice turns out  to be $\mathcal{R}_{0}=10^{-6}$, for it satisfies the average radius of the outer edge of a Schwarzschild black hole accretion disk at $r \approx 70r_g$ by setting the innermost and outermost stable circular orbits, according to Eq. (\ref{r0}), at  $x_c=6$ and $x_c=143.45$ \cite{2013A&A551A4P}.

A similar analysis, carried out for the Kerr metric in modified gravity with the metric elements,
\begin{alignat}{2}\label{kerrmetric}
& g_{tt}=-\alpha^2=-\frac{(\Delta_r-a^2)}{\Xi^2r^2}\ ;\  &\qquad\text{} g_{rr}=\frac{r^2}{\Delta_r},\notag \\
& g_{t\phi}=-\frac{2a}{\Xi^2r^2}(r^2+a^2-\Delta_r)\ ; \notag \\ 
&\qquad\text{} g_{\phi\phi}=\frac{(r^2+a^2)^2-\Delta_ra^2}{\Xi^2r^2}
\end{alignat}
with
\begin{eqnarray}
\Delta_r=(r^2+a^2)(1-\frac{R_0}{12}r^2)-2r_gr,\\
\Xi=1+\frac{R_0}{12}a^2,
\end{eqnarray}
demands $\mathcal{R}_{0}$ (upon setting $1/g_{rr}=0$) to satisfy
\begin{equation}\label{kerreq}
(x^2+\frac{a^2}{r_g^2})\left(1-\frac{\mathcal{R}_0x^2}{12}\right)-2x=0.
\end{equation}
While Eq.($\ref{kerreq}$) yields the range for $\mathcal{R}_{0} \in [-0.3,0.6]$, a new range  for $\mathcal{R}_{0} \in (0,0.6]$ emerges if a Kerr black hole with two event horizons and one cosmological horizon is demanded. However, once again, upon demanding $v_{eff}=0$, $dV_{eff}/dr=0$, and $d^{2}V_{eff}/d^{2}r\geq 0$ along with the appropriate temperature and luminosity profiles mentioned above, we find the stable circular orbits can exist only for $\mathcal{R}_{0} \in [-1.2\times 10^{-3}, 6.67\times 10^{-4}]$. To maintain the consistency in our numerical plots, we again choose $\mathcal{R}_{0}=10^{-6}$, for it satisfies the the average radius of the outer edge of a Kerr black hole accretion disk at $r \approx 16r_g$ by setting the innermost and outermost stable circular orbits at  $x_c=1.4545$ and $x_c=143.45$, respectively. It should be noted here that temperature profiles used for both classes of accretion disks remain the same as long as we choose $\mathcal{R}_{0}\approx 10^{-6}$ \cite{2013A&A551A4P}.

The general expression of vorticity generated in modified gravity is 
\begin{align}\label{generic2}
\vec{\Omega}(r)=-(1+\mathcal{R}_0)\frac{\chi k_b c r\zeta}{q}\frac{1}{\sqrt{g_{rr}g_{\phi\phi}}}\partial_r(\Gamma_m^{-1})\left(\partial_rT\right) \Delta t \ \hat{\theta},
\end{align}
with $\Gamma_m$ and $\lambda f_{m}(R) =\mathcal{R}_0$ denoting the modified Lorentz factor and  the nonminimal coupling of plasma to $f(R)$ gravity, respectively. Similarly to Eq.(\ref{genericsch}), we have an analytical expression for modified Schwarzschild spacetime
\begin{align}\label{mgenericsch}
\vec{\Omega}(x)=-(1+\mathcal{R}_0)\frac{3\zeta k_b\pi\Gamma_m\alpha_m\chi}{q\sqrt{x}r_g}\partial_xT \ \hat{\theta},
\end{align}
where $\alpha_m$ is associated with modified Schwarzschild metric component $g_{tt}$. 
\begin{figure}
	\includegraphics[width=0.90\linewidth]{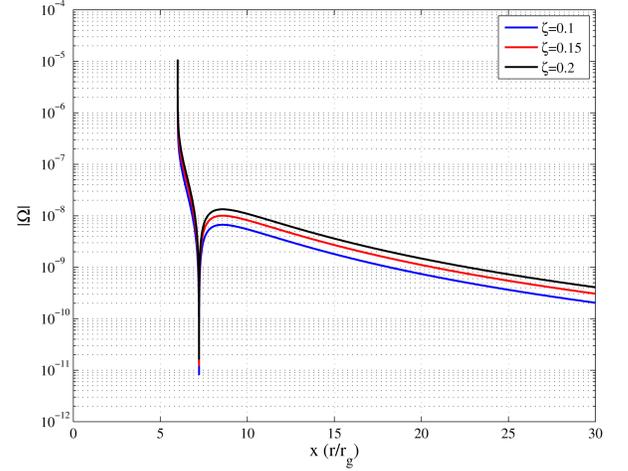}
	\caption{Vorticity $\Omega$ (in the units of Gauss) for Schwarzschild black hole with different $\zeta$ values in modified gravity within $6.0 r_g$ and $30 r_g$ for  $M=14.8M_{\odot}$.}
	\label{fig:msch_mod}
\end{figure}
\begin{figure}
	\includegraphics[width=0.90\linewidth]{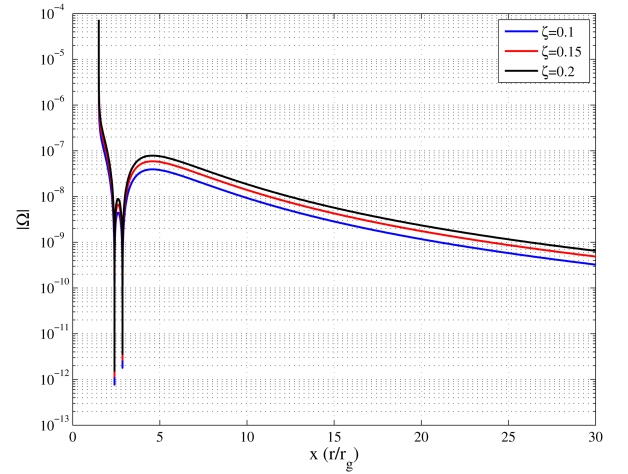}
	\caption{Vorticity $\Omega$ (in the units of Gauss) for Kerr black hole with different $\zeta$ values in modified gravity within $1.5 r_g$ and $30 r_g$ for  $M=14.8M_{\odot}$ with an accretion rate  $\dot{M}=0.472\times 10^{19}g/s$ and a=0.99.}
	\label{fig:mkerr_mod}
\end{figure}

Figures $\ref{fig:msch_mod}$ and \ref{fig:mkerr_mod} show the plot of this generalized vorticity ${|\Omega|}$ (in the disk) as a function of the distance $x=r/r_g$ in the gravitational field of a black hole of mass $M=14.8M_{\odot}$ and $\mathcal{R}_0=10^{-6}$. The kinklike behavior of  $|\Omega|$ vs $x=r/r_g$, as earlier, originates in the vanishing gradients of temperature and the Lorentz factor. Different values of $\zeta$, used in the Figs. $(\ref{fig:msch_mod})$ and (\ref{fig:mkerr_mod}), capture the influence of the modified gravity on toroidal temperature fluctuations. Here, it should be emphasized that vorticity magnitude profile in the modified gravity can be drastically different for arbitrary functions of $f(R)$, which signifies how the matter is coupled to its background spacetime.

It must be noted that, as maximum temperature obtained in $f(R)$ models turns out to be lower than Cygnus X-1 observation values, the probability of the existence of an accretion disk in the $f(R)$ Schwarzschild black hole becomes very slim \cite{gou2011}. Still we present the result here for $f(R)$ Schwarzschild black hole as an analytical example.

\subsection{Relative strength of relativistic and baroclinic drive}

The generalized magnetofluid dynamics, derived for a barotropic equation of state, will have only the relativistic vorticity drives.  We would now like to compare the relative magnitudes of the  relativistic,  and possibly baroclinic, vorticity sources. At first glance, it is evident (from the metric) that the relativistic source will dominate the baroclinic source as we approach the event horizon. But to make a comparison between the relativistic and baroclinic drives, let us construct a simple baroclinic drive of the form  $\nabla T\times\nabla\sigma\approx (\epsilon/r)g_{\phi\phi}^{-1/2}\partial_{\phi}(k_bT)$, where  $\epsilon$ is a measure of  the departure from strict barotropic behavior. The preceding  approximation introduces a smaller toroidal temperature variation that quantifies the nonbarotropic component. 

\begin{figure}
	\includegraphics[width=0.90\linewidth]{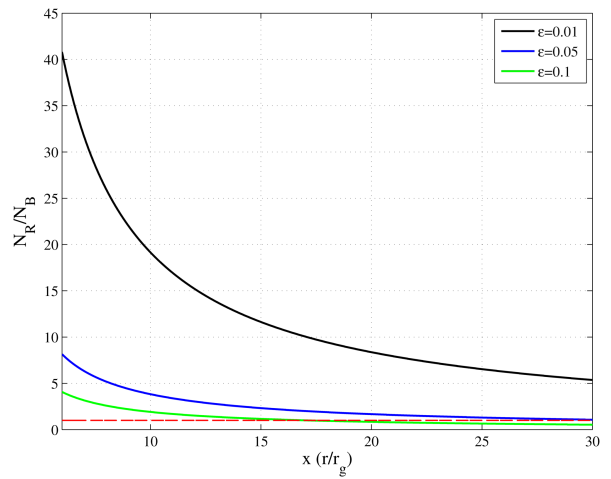}
	\caption{Ratio between relativistic and baroclinic drive in Schwarzschild black hole with $\epsilon$=0.01 (black), 0.05(green) and 0.1(blue).The red line represents the ratio of unity.}
	\label{fig:ratio_SBH}
\end{figure}

\begin{figure}
	\includegraphics[width=0.90\linewidth]{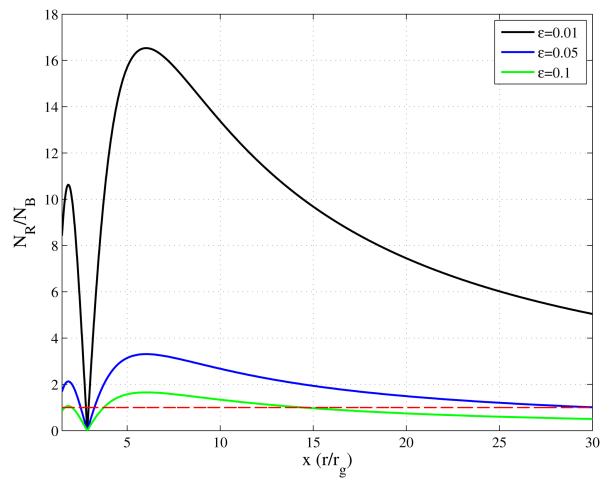}
	\caption{Ratio between the relativistic and baroclinic drive in the Kerr black hole with with $\epsilon$=0.01 (black), 0.05(green) and 0.1(blue). The red line represents the ratio of unity.}
	\label{fig:ratio_KBH}
\end{figure}

In Figs. (\ref{fig:ratio_SBH}) and (\ref{fig:ratio_KBH}), we compare the relative strength of the relativistic and baroclinic drive for black hole accretion disks in Schwarzschild and Kerr geometries. We plot the relative magnitude as a function of the radial distance, $x=r/r_g$;  both plots start from their respective isco. In both cases, the relativistic drive becomes dominant as we approach the inner most stable orbit. For smaller values of $\epsilon$, we see the magnitude of the relativistic drive keeps increasing as the departure from barotropic fluid is minimal. Also, the dashed line shows the ratio to be unity and from both figures we see that the relativistic drive remains dominant, for smaller values of $\epsilon$; the dominance continues to longer distances from innermost stable circular orbit. In both cases, the Lorentz factor plays an important role in determining their relative strength. Thus, the sudden dip in the relative strength profile in Kerr metric can be attributed to the vanishing gradient of the corresponding Lorentz factor inherent in the relativistic drive. Moreover, the relative magnitude in the Kerr black hole is less than that in Schwarzschild black hole because of the significant energy loss of plasmas in Kerr space-time due to gravitational radiation reaction as it approaches the nonstable orbits. The jump in relative intensity from $\epsilon= 0.01$ to  $\epsilon= 0.05$ remains significant in both  geometries.

\section{Conclusion}
We have explored in this paper the possibility of  generating  what may be called ``generalized vorticity" in accretion disks surrounding compact gravitating objects, in particular, the Kerr and Schwarzschild black holes. The accretion disc plasma is coupled nonminimally [via $f(R)$ gravity] to the surrounding specified space-time. Although vorticity can be generated by the well-known baroclinic mechanism ( nonzero $\nabla T\times\nabla\sigma$), we have concentrated, in this paper, on exploring  what  are classed as relativistic drives stemming from space-time distortions caused by special  as well as general relativistic effects. We find that the Kerr geometry, due to the intrinsic rotation, is a more efficient vorticity generator as compared to the Schwarzschild counterpart. We also observe a slight increase in  vorticity generation even if an extremely weak $f(R)$ (nonminimal) coupling, $f(R)=R_0=constant$, is turned on, which implies that a noticeable change in vorticity generation will be observed in strong $f(R)$ coupling. Physically, the increase in efficiency is directly related to the amount of deviation from the circular orbits (of the plasma particles) caused by the distorted geometry.

We also compared the efficiency of the relativistic drive  with a model  baroclinic drive. The strength of the baroclinic drive is given in terms of a parameter  $\epsilon$, that measures the departure from strict barotropy. For most accretion disk plasmas in quasiequilibrium, $\epsilon$ is expected to be small. We find that, for reasonable values of $\epsilon$, the relativistic drive is dominant for the disk regions nearer to the compact object; as the distance increases, and for relatively larger $\epsilon$, the barcolinic term becomes comparable to the relativistic drive. This vorticity growth occurs in orbital time scale, similar to MRI instability, which later can amplify through several dynamo mechanisms \cite{marklund2005general,reinhardt1975note,Klahr:2004ip}. 

Unlike GR-MHD, our formalism can be extended to multifluid species with each species obeying its  its own vortical dynamics\cite{mckinney2012general,penna2010simulations}. In addition, the induction equation in this formalism has source terms for vorticity generation; these sources can catapult the accretion disk to a state of finite vorticity from one with no vorticity (electromagnetic and hydrodynamic). This formalism can also be studied in the context of vortex generation in the protoplanetary disks near a supermassive star where gravity plays a dominant role. Hyrdodynamic simulations indicate these proto-planetary disks to be inherently baroclinic due to negative radial entropy gradient\cite{0004-637X-765-2-115,Klahr:2002sm,Klahr:2004ip}. The turbulence caused by the baroclinic instabilities is found to be suitable for angular momentum transport and vortex formation in the disk which are suggested to lead to planet formation. Our model, if applied to the evolution of the protoplanetary disk, will provide additional sources for angular momentum transport and vortex formation even in a barotropic disk.

Regardless of the details, the curvature drive (in minimal as well as in nonminimal gravity) will always generate ``generalized" vorticity  which, either by itself or after amplification through a generalized dynamo mechanism, can provide a mechanism for angular momentum transport. These amplified vortical fields can also collimate the jets emanating from the disk where field lines corotate with the disk, by the flux freezing theorem. The plasma leaving the disk can drag the field lines due to its large conduction coefficient which wraps the field lines around the rotation axis. The field lines then exert a radial force which can compress the jet of plasma leading to jet collimation \cite{vietri2008foundations}.

\section*{Acknowledgments}
Authors thank David J. Stark and Manasvi Lingam for helpful discussions. C.B and S.M.M's research was supported by the US DOE Grant No. DE-FG02-04ER-54742.  


\begin{appendix}
	
	\section{3+1 Dynamics of GravitoMagnetofluid }\label{magnetofluiddynamics}

	The approach chosen for the 3+1 splitting selects a family of foliated fiducial 3-dimensional hypersurfaces (slices of simultaneity) $\Sigma_{t}$ labelled by a parameter $t = constant$ in terms of a time function on the manifold. Furthermore, we let $t^{\mu}$ be a timeline vector whose integral curves intersect each leaf $\Sigma_{t}$ of the foliation precisely once and which is normalized such that $t^{\mu}\nabla_{\mu}t = 1$. This $t^{\mu}$ is the `evolution vector field' along the orbits of which different points on all  $\Sigma_{t} \equiv \Sigma$ can be identified. This allows us to write all space-time fields in terms of $t$-dependent components defined on the spatial manifold $\Sigma_{t}$. Lie derivatives of space-time field along $t^{\mu}$ are identified with ``time derivatives" of the spatial fields since Lie derivatives reduce to partial time derivative for an adapted coordinate system $t^{\mu}=(1,0,0,0)$.
	
	Moreover, since we are using the Lorentzian signature, the vector field $t^{\mu}$ is required to be future directed. Let us decompose $t^{\mu}$ into normal and tangential parts with respect to $\Sigma_{t}$ by defining the lapse function $\alpha$ and the shift vector $\beta^{\mu}$ as $t^{\mu}=\alpha n^{\mu}+\beta^{\mu}$ with $\beta^{\mu}n_{\mu} = 0$, where  $n^{\mu}$ is the future directed unit normal vector field to the hypersurfaces $\Sigma_{t}$. More precisely, the natural timelike covector $n_{\mu}=(-\alpha,0,0,0)=-\alpha\nabla_{\mu}t$  is defined to obtain $n^{\mu}=({1}/{\alpha},-\beta^{\mu}/\alpha)$ which satisfy the normalization condition $n^{\mu}n_{\mu}=-1$. Then, the space-time metric $g_{\mu\nu}$ induces a spatial metric $\gamma_{\mu\nu}$ by the formula $\gamma_{\mu\nu}=g_{\mu\nu}+n_{\mu}n_{\nu}$. Finally, the 3+1 decomposition is usually carried out with the projection operator $\gamma^{\mu}\ _{\nu}=\delta^{\mu} \ _ {\nu}+n^{\mu}n_{\nu}$, which satisfies the condition $n^{\mu}\gamma_{\mu\nu}=0$. Also, the acceleration is defined as $a_{\mu}=n^{\nu}\nabla_{\nu}n_{\mu}$. \\
	
	Now, with the above foliation of space-time, the space-time metric  takes the following canonical form \cite{MTW}
	\begin{equation}\label{canonicalmetric}
	ds^2=-\alpha^2dt^2+\gamma_{ij}(dx^i+\beta^i dt)(dx^j+\beta^j dt),
	\end{equation}
	and it immediately follows that, with respect to an Eulerian observer, the Lorentz factor turns out to be
	\begin{equation}\label{lorentzfactor}
	\Gamma=\left[\alpha^2-\gamma_{ij}(\beta^{i}\beta^{j}+2\beta^{i}v^{j}+v^{i}v^{j})\right]^{-1/2},
	\end{equation}
	satisfying $d\tau = dt/\Gamma$, where $v^i$ is the $i$th component of fluid velocity $\vec{v}=d\vec{x}/dt$.  Then the decomposition for the 4-velocity is \cite{asenjo2013generating}
	\begin{equation}\label{fourvelocity}
	U^{\mu}=\alpha \Gamma n^{\mu}+\Gamma\gamma^{\mu} \ _{\nu}v^{\nu},
	\end{equation}
	with $n_{\mu}U^{\mu}=-\alpha \Gamma$.

	Now, since our unified anti-symmetric field tensor $\mathcal{M}^{\mu\nu}$ is constructed from the antisymmetric tensors $F^{\mu\nu}$ and $D^{\mu\nu}$, we apply the ADM formalism of electrodynamics presented in \cite{thorne1982electrodynamics,thorne1986black,MTW,Wald} to define the generalized electric and magnetic field, respectively, as 
	\begin{alignat}{2}\label{generalemdef}
	\xi^{\mu}=n_{\nu}\mathcal{M}^{\mu\nu}\ ;\  &\qquad\text{}\qquad X^{\mu}=\frac{1}{2}n_{\rho}\epsilon^{\rho\mu\sigma\tau}\mathcal{M}_{\sigma\tau},
	\end{alignat}
	and thus express the unified field tensor
	\begin{equation}\label{generalfieldtensordef}
	\mathcal{M}^{\mu\nu}=n^{\mu}\xi^{\nu} - n^{\nu}\xi^{\mu}-\epsilon^{\mu\nu\rho\sigma}X_{\rho}n_{\sigma}.
	\end{equation}
	We remind the reader that the generalized magnetic field and the generalized vorticity are essentially synonymous. Using the definition of the unified field tensor $\mathcal{M}^{\mu\nu}$,  the expressions of 3D generalized electric and magnetic fields turn out to be
	
	\begin{align}\label{generalefield}
	\vec{\xi}= \ & \vec{E}-\frac{m}{q}(1+\lambda f_{m}(R)-\lambda RF_{m}(R))\vec{\nabla}(\alpha \mathcal{G}\Gamma)\notag \\
	&-\frac{m}{q}\lambda F_{m}(R)\vec{\nabla}(\alpha \mathcal{G}R\Gamma)\notag \\
	& -\frac{m}{q}(1+\lambda f_{m}(R))\left[2\underline{\underline{{\sigma}}}\cdot(\mathcal{G}\Gamma\vec{v})+\frac{2}{3} \Theta\mathcal{G}\Gamma\vec{v}\right]\notag \\
	& -\frac{m}{q\alpha}(1+\lambda f_{m}(R)-\lambda RF_{m}(R))\left(\mathcal{L}_t(\mathcal{G}\Gamma\vec{v}) -\mathcal{L}_{\vec{\beta}}(\mathcal{G}\Gamma\vec{v})\right)\notag  \\
	& -\frac{m}{q\alpha}\lambda F_{m}(R)\left(\mathcal{L}_t(\mathcal{G}R \Gamma\vec{v}) -\mathcal{L}_{\vec{\beta}}(\mathcal{G}R \Gamma\vec{v})\right);
	\end{align}
	\begin{align}\label{generalbfield}
	\vec{X}=\vec{B}+\frac{m}{q}(1+\lambda f_{m}(R)-\lambda RF_{m}(R)) \vec{\nabla}\times(\mathcal{G}\Gamma\vec{v})\notag \\
	+\lambda F_{m}(R) \frac{m}{q}\vec{\nabla}\times(R\mathcal{G}\Gamma\vec{v}),
	\end{align}
	where $\underline{\underline{{\sigma}}} = \sigma_{\mu}^{\nu}$ and $\Theta$ are, respectively,   the shear and expansion of the congruence, defined as $\sigma_{\alpha\beta} = \gamma_{\alpha}^{\mu}\gamma_{\beta}^{\nu}\nabla_{(\mu} n_{\nu)}- \frac{1}{3}\theta \gamma_{\mu\nu}$ and $\Theta =\nabla_{\mu}n^{\mu} $. We have also used the relation $\nabla_{\mu} n_{\nu} = -a_{\nu}n_{\mu} + \sigma_{\alpha\beta} + \frac{1}{3}\theta \gamma_{\mu\nu}$ to derive (\ref{generalefield}).

\end{appendix}

\bibliographystyle{unsrt}
\bibliography{ref}

\end{document}